\documentclass[showkeys]{revtex4-2}

\usepackage{multirow}
\usepackage{amsmath,amsfonts}
\usepackage{amssymb,latexsym}
\usepackage{color}
\usepackage{graphicx}
\usepackage{mathrsfs}
\usepackage[normalem]{ulem} % for striking text

\newtheorem{prop}{Proposition}

\newcommand{\beprop}{\begin{prop}}
\newcommand{\enprop}{\end{prop}}
\newcommand{\bprf}{\begin{proof}} 
\newcommand{\eprf}{\end{proof}\qed}

\definecolor{hervecolor}{rgb}{0.8,0,0.7}

\newcommand{\ket}[1]{|\kern.3ex#1\kern.3ex\rangle}
\newcommand{\bra}[1]{\langle\kern.3ex #1 \kern.3ex|}
\newcommand{\scalar}[2]{\langle\kern.3ex #1 \kern.3ex|\kern.3ex#2\kern.3ex\rangle}

\newcommand{\ii}{\mathsf{i}}

\newcommand{\uE}{\mathrm{E}}

\begin{document}

\title{Does canonical quantization lead to GKSL dynamics?}
\author{T.\ Koide}
\email{tomoikoide@gmail.com,koide@if.ufrj.br}
\affiliation{Instituto de F\'{\i}sica, Universidade Federal do Rio de Janeiro, C.P. 68528,
21941-972, Rio de Janeiro, RJ, Brazil}
\author{F.\ Nicacio}
\email{nicacio@if.ufrj.br}
\affiliation{Instituto de F\'{\i}sica, Universidade Federal do Rio de Janeiro, C.P. 68528,
21941-972, Rio de Janeiro, RJ, Brazil}

\begin{abstract}
We introduce a generalized classical model of Brownian motion for describing thermal relaxation processes 
which is thermodynamically consistent.
Applying the canonical quantization to this model, a quantum equation for the density operator is obtained.
This equation has a thermal equilibrium state as its stationary solution, 
but the time evolution is not necessarily a 
Completely Positive and Trace-Preserving (CPTP) map. 
In the application to the harmonic oscillator potential, however, 
the requirement of the CPTP map is shown to be satisfied 
by choosing parameters appropriately and then our equation reproduces 
a Gorini-Kossakowski-Sudarshan-Lindblad (GKSL) equation satisfying the detailed balance condition.
This result suggests a quantum-classical correspondence in thermal relaxation processes and 
will provide a new insight to the study of decoherence.
\end{abstract}


\maketitle

\section{Introduction}

The accelerating development of experimental methods enables us to access individual thermal random processes in a small fluctuating system.
Standard thermodynamics is not directly applicable to this system because the macroscopic and microscopic time scales are not necessarily clearly separated and the effect of fluctuations is important.
There is no established method for describing such systems, but it is known that Brownian motion can be used to describe them 
consistently with thermodynamics.
Such a framework is called stochastic energetics (stochastic thermodynamics) \cite{sekimoto,seifert,peliti}.

It should be understood that classical stochastic models do not consider quantum fluctuation and hence 
are not applicable to extremely microscopic systems.
Regarding this limitation, there are various studies to derive quantum dissipative equations from 
either the Heisenberg equations of motion or the Liouville-von Neumann equations,
introducing systematic coarse-graining with, for example, projection operator techniques \cite{zwanzig-book,breuer}. 
However, the derived equations often do not satisfy the (complete) 
positivity of probability and its conservation (preservation of trace of a density operator) at the same time.
The linear time evolutions satisfying these requisites are called Completely Positive and Trace-Preserving (CPTP) maps \cite{breuer}. 
Such a map, when the Markovianity is assumed, 
leads to the well-known Gorini-Kossakowski-Sudarshan-Lindblad (GKSL) equation, 
which is used to describe quantum thermal relaxation processes \cite{breuer}.
Indeed, by applying this to the harmonic potential and imposing the detailed balance conditions 
to the parameters, the behavior of the GKSL equation is  
compatible with the laws of thermodynamics \cite{spohn,alicki,davies,spohn2,pusz}.

In standard applications, quantum dynamics is obtained from the corresponding classical equation  
through the canonical quantization \cite{sakuray}. 
Therefore, it is natural to ask whether a GKSL equation can be derived from Brownian motion provided 
the canonical quantization recipe.
This question is investigated by Oliveira in Refs.\ \cite{oliveira1,oliveira2}, 
where he considers the Kramers (Fokker-Planck) equation for a classical particle interacting with a heat bath.
Regarding it as the classical correspondence of the density operator equation (master equation) in quantum physics, his quantum dissipative equation is obtained by the canonical quantization, 
which however does not reproduce a GKSL equation.
Prior to Oliveira, Burzlaff studies conditions for the Liouville equation to be mapped into 
a classical counterpart of the GKSL equation \cite{burzlaff}.
Such a mapping is not always found and this approach is
not applicable to thermal relaxation processes which are
not described by the Liouville equation.
The applicability of the canonical quantization is thus not known in the derivation of the GKSL equation. 

In this paper, we show that the GKSL equation is obtained from 
a generalized Kramers equation, at least, 
for the harmonic oscillator Hamiltonian.
There are two differences comparing to the approach by Oliveira.
First, the interaction with heat baths affects not only the equation of momentum but also that of position of Brownian motion.
Although this seems to be fanciful in standard stochastic processes, we should recall that position and momentum are not distinguished by the symplectic structure of the phase space. 
Moreover, in quantum physics,  
it is known that the equation of position is modified by random noises induced 
by quantum fluctuation 
in the derivation of the Schr\"{o}dinger equation based on Brownian motion
which is studied by Nelson and Yasue \cite{nelson,yasue,zambrini,koide_review}, for instance.
Indeed, we show that this generalized model of Brownian motion is consistent with the 
thermodynamical laws. 
Second, 
supported by the non-uniqueness of the symmetrization among products of position and momentum operators in the quantization procedure, 
we adopt a different symmetrization procedure, while preserving the standard mapping between the commutator 
and the Poisson bracket of canonically conjugated variables of position and momentum.
This symmetrization is inspired by the relation found in the Fokker-Planck operator considered 
in Refs.\ \cite{koide_rbm2,koide_fp1,koide_fp2,kurchan}.

This paper is organized as follows.
The generalized model of Brownian motion is developed in Sec.\ \ref{sec:cla_model}.
The canonical quantization of this classical model is studied in Sec.\ \ref{sec:cq}. 
When we apply this quantum model to the harmonic oscillator potential, the GKSL equation is reproduced. 
Section \ref{sec:conc} is devoted to the concluding remarks.

\section{Generalized thermal relaxation model} \label{sec:cla_model}

Brownian motion is a standard model employed to describe classical 
thermal relaxation process \cite{sekimoto,seifert,peliti}, 
where the temporal evolution described by canonical equations in the absence of an environment 
is modified by introducing two primary effects: a friction force and a random thermal force due to the interaction 
with a heat bath at temperature $T$.
For a $D+1$ dimensional single-particle system of mass $m$, 
the stochastic differential equations for position $\widetilde{\bf q}_t$ and ``momentum" 
$\widetilde{\bf p}_t$ are typically given by 
\begin{eqnarray}
d\widetilde{\bf q}_{t} &=&\frac{\widetilde{\bf p}_t}{m} \label{eqn:1}\, ,\\
d\widetilde{\bf p}_{t} &=& - \frac{\partial V(\widetilde{\bf q}_t)}{\partial \widetilde{\bf q}_{t}}dt  
- \gamma \frac{\widetilde{\bf p}_t}{m} dt + \sqrt{2\nu} d\widetilde{\bf B}_t \label{eqn:2}\, ,
\end{eqnarray}
where $V(\widetilde{\bf q}_t)$ denotes an external potential and the symbol 
``$\,\widetilde{\bullet}\,$'' is conveniently introduced to denote stochastic variables.
In the above equations, the first terms on the right-hand sides are recognized as the canonical equations, 
while the subsequent terms in the second equation represents  
the friction force and the random thermal force  induced by the interaction with a heat bath.
The parameters $\gamma$ and $\nu$ quantify the strengths of these forces and, 
to describe a thermal relaxation process, these must be related by $\nu = \gamma k_B T$, 
which is the expression of the fluctuation-dissipation theorem (Einstein relation). 
The random thermal influences are described in terms of the Wiener process $\widetilde{\bf B}_t$, 
which is described later on.

The aforementioned discussion will be now extended to a $D+1$ dimensional system composed of $N$ particles. 
In the absence of environmental interaction, the Hamiltonian describing the closed system is
\begin{eqnarray}
H(\{ \widetilde{\bf q}_t , \widetilde{\bf p}_t \},  \vec{\lambda}_t) \, ,
\end{eqnarray}
where $\{\widetilde{\bf q}_t , \widetilde{\bf p}_t \} = ( \widetilde{\bf q}_{(1)t} , \widetilde{\bf p}_{(1)t}), \cdots,  (\widetilde{\bf q}_{(N)t}, \widetilde{\bf p}_{(N)t})$. 
The Hamiltonian depends on a set of external parameters $ \vec{\lambda_t} =(\lambda^{a}_t, \lambda^{b}_t, \cdots)$ 
through of which we control, for example, the form of an external (confinement) potential of the $N$-particles.
The Hamiltonian is arbitrary, possibly containing inter-particle interactions.
Moreover, a relativistic thermal relaxation observed in the rest frame of heat baths 
is described by choosing a relativistic Hamiltonian and the result is consistent with relativistic stochastic energetics \cite{koide_rbm1,koide_rbm2,deffer}.

We introduce $N$-independent heat baths of temperature $T_i$.
Our derivation relies primarily on the assumption that the dynamics of this system can be captured by the subsequent stochastic differential equations for 
$i=1,\dots, N$,
\begin{eqnarray}
d\widetilde{\bf q}_{(i)t} &=& \left( \frac{\partial H(\{ \widetilde{\bf q}_t , \widetilde{\bf p}_t \},  \vec{\lambda}_t) }{\partial \widetilde{\bf p}_{(i)t}} 
- \gamma_{q_i}  \frac{\partial H(\{ \widetilde{\bf q}_t , \widetilde{\bf p}_t \},  \vec{\lambda}_t) }{\partial \widetilde{\bf q}_{(i)t}}  \right) dt 
+ \sqrt{\frac{2\gamma_{q_i} }{\beta_i}} d\widetilde{\bf B}_{q(i)t}
\, , \label{eqn:eq_x}\\
d\widetilde{\bf p}_{(i)t} &=& \left( - \frac{\partial H(\{ \widetilde{\bf q}_t , \widetilde{\bf p}_t \},  \vec{\lambda}_t) }{\partial \widetilde{\bf q}_{(i)t}}  
- \gamma_{p_i}  \frac{\partial H(\{ \widetilde{\bf q}_t , \widetilde{\bf p}_t \},  \vec{\lambda}_t) }{\partial \widetilde{\bf p}_{(i)t}} \right)  dt 
+ \sqrt{\frac{2\gamma_{p_i} }{\beta_i}} d\widetilde{\bf B}_{p(i)t}  \label{eqn:eq_p}\, ,
\end{eqnarray}
where $dt>0$, $\beta_i = 1/(k_B T_i)$ for the Boltzmann constant $k_B$, and $\gamma_{q_i}$ and $\gamma_{p_i}$ are real constants. 
The $i$-th particle directly interacts with the $i$-th heat bath, 
while the remaining particles are coupled with the same bath via friction forces, due to inter-particle interactions.
It can be readily verified that these equations reduce to Eqs.\ (\ref{eqn:1}) and (\ref{eqn:2}) 
when employing an appropriate Hamiltonian for a single particle and setting $\gamma_{q_i} = 0$. 
In this context, our model represents a generalization of the standard Brownian motion \cite{sekimoto,seifert,peliti} 
due to the interaction of the baths with the stochastic positions $d\widetilde{\bf q}_{(i)t}$, 
since the second and third terms on the right-hand side of Eq.\ (\ref{eqn:eq_x}) 
are absent in the standard description, which is recovered by setting $\gamma_{q_i} = 0$.

As in standard Brownian motion, the random thermal forces or noise terms are 
expressed in terms of the Wiener processes $\widetilde{\bf B}_{q(i)t}$ and $\widetilde{\bf B}_{p(i)t}$. 
The inclinations are defined by 
$d\widetilde{\bf B}_{q(i)t} = \widetilde{\bf B}_{q(i)t+dt} - \widetilde{\bf B}_{q(i)t}$ and 
$d\widetilde{\bf B}_{p(i)t} = \widetilde{\bf B}_{p(i)t+dt} - \widetilde{\bf B}_{p(i)t}$, and 
these satisfy the following correlation properties \cite{gardiner}:
\begin{eqnarray}
&& \uE \! \left[  d\widetilde{B}_{q(i)t} \right] = \uE \! \left[  d\widetilde{B}_{p(i)t} \right] = 0 \, , \\
&& \uE \! \left[  d\widetilde{B}_{q(i)t} d\widetilde{B}_{q(j)t^\prime} \right] = \uE \! \left[  d\widetilde{B}_{p(i)t} d\widetilde{B}_{p(j)t^\prime} \right] =  dt \, \delta_{ij}\delta_{t,t^\prime}  \, ,\\
&& \uE \!  \left[  d\widetilde{B}_{q(i)t} d\widetilde{B}_{p(j)t^\prime} \right] = 0 \, ,
\end{eqnarray}
where $\uE [\, \bullet \,]$ denotes the ensemble average for the Wiener process.
It should be emphasized that $({\bf q}_{(i)t}, {\bf p}_{(i)t})$ form canonical pairs for $H(\{ {\bf q}_t , {\bf p}_t \},  \vec{\lambda}_t)$, 
but are not necessarily canonical variables under the influence of the heat baths, because we do not define the Lagrangian for such a system.

Introducing a normalized phase space distribution $f(\{ {\bf q}, {\bf p} \},t)$, its temporal evolution
is obtained by using Ito's lemma \cite{gardiner} 
and the stochastic differential equations (\ref{eqn:eq_x}) and (\ref{eqn:eq_p}): 
\begin{eqnarray}
\partial_t f(\{ {\bf q}, {\bf p} \},t)
&=& -\{  f(\{ {\bf q}, {\bf p} \},t) , H\}_{\text{PB}}  \nonumber \\ 
&&+  \sum_{i=1}^N \sum_{\alpha=1}^D \frac{\gamma_{p_i} }{\beta_i} \{ e^{-\beta_i H}\{ e^{\beta_i H} f(\{ {\bf q}, {\bf p} \},t) , \, q^{\alpha}_{(i)} \}_{\text{PB}}, \, q^{\alpha}_{(i)} \}_{\text{PB}} 
\nonumber \\
&& +  \sum_{i=1}^N \sum_{\alpha=1}^D \frac{\gamma_{q_i}}{\beta_i} \{ e^{-\beta_i H}\{  e^{\beta_i H} f(\{ {\bf q}, {\bf p} \},t), \, p^{\alpha}_{(i)} \}_{\text{PB}}, \, p^{\alpha}_{(i)} \}_{\text{PB}} \, , 
\label{eqn:kramers}
\end{eqnarray}
where, for arbitrary functions $g$ and $h$, the Poisson bracket is defined by
\begin{eqnarray}
\{g,h\}_{\text{PB}} = \sum_{i=1}^N \sum_{\alpha=1}^D \left( \frac{\partial g}{\partial q^\alpha_{(i)}}\frac{\partial h}{\partial p^\alpha_{(i)}}
- \frac{\partial g}{\partial p^\alpha_{(i)}}\frac{\partial h}{\partial q^\alpha_{(i)}} \right) \, .\label{eqn:pb}
\end{eqnarray}
See Appendix \ref{app:0} for details. 
The bath-position couplings considered in the stochastic equation (\ref{eqn:eq_x}) appears 
in the third term on the right hand side of Eq.\ (\ref{eqn:kramers}), 
thus this is called here the generalized Kramers equation. Like the standard version, this generalization also drives any initial distribution asymptotically towards the thermal equilibrium state, as shown in Appendix \ref{app:1}.

To unveil the thermodynamical structure, we define the absorbed heat from the heat baths as the works done by 
these baths \cite{sekimoto}, 
\begin{eqnarray}\label{eqn:heat}
d\widetilde{Q}^c_t 
= 
\sum_{i=1}^N d\tilde{Q}^c_{(i)t}  \, ,
\end{eqnarray}
where
\begin{eqnarray} \label{eqn:heat_sum}
d\widetilde{Q}^c_{(i)t} 
&=& 
 \sum_{\alpha =1}^D
\left( - \gamma_{p_i} (t) \frac{\partial H(\{ \widetilde{\bf q}_t , \widetilde{\bf p}_t \},  \vec{\lambda}_t) }{\partial \widetilde{p}^\alpha_{(i)t}}  dt 
+ \sqrt{\frac{2\gamma_{p_i} (t)}{\beta_i}} d\widetilde{B}^\alpha_{p(i)t} \right) \circ d\widetilde{q}^{\alpha}_{(i)t} 
\nonumber \\
&&
-
\sum_{\alpha =1}^D
\left( - \gamma_{q_i} (t) \frac{\partial H(\{ \widetilde{\bf q}_t , \widetilde{\bf p}_t \},  \vec{\lambda}_t) }{\partial \widetilde{q}^{\alpha}_{(i)t}}  dt + \sqrt{\frac{2\gamma_{q_i} (t)}{\beta_i}} d\widetilde{B}^{\alpha}_{q(i)t} \right) \circ d\widetilde{p}^{\alpha}_{(i)t} \, .
\end{eqnarray}
Here the symbol ``$\circ$'' 
indicates the Stratonovich definition for
the product of stochastic quantities \cite{gardiner}.
The second term on the right-hand side does not exist in stochastic energetics and is induced 
in our description by interactions with heat baths in Eq.\ (\ref{eqn:eq_x}). 
See Appendix \ref{app:2} for more details.

The expectation value of the Hamiltonian defines the energy of the system through 
\begin{eqnarray}
E^c_t &=&  \int d\Gamma_0 f_{0} (\{ {\bf q}_0, {\bf p}_0 \}) \, \uE [H(\{ \widetilde{\bf q}_t , \widetilde{\bf p}_t \}, \vec{\lambda}_t)] \, ,
\end{eqnarray}
where 
$\{ {\bf q}_0, {\bf p}_0 \}$ are the position and momentum at an initial time,  
$d\Gamma_0$ is the corresponding phase space volume, 
and $f_{0} (\{ {\bf q}_0, {\bf p}_0 \})$ is the initial probability distribution. 
The change of energy
induced by the control parameters $\vec{\lambda}_t$ is interpreted as the work done to the system,
\begin{eqnarray}
dW^c_t &=& \int d\Gamma_0 \, f_{0} (\{ {\bf q}_0, {\bf p}_0 \}) \, \uE \!\left[ \sum_{j} d\lambda^j_t  \frac{\partial H(\{ \widetilde{\bf q}_t , \widetilde{\bf p}_t \},  \vec{\lambda}_t)}{\partial \lambda^j_t}   \right] \, ,
\end{eqnarray}
while, from Eq.(\ref{eqn:heat}), 
\begin{eqnarray}
dQ^c_t &=& \int d\Gamma_0 f_{0} (\{ {\bf q}_0, {\bf p}_0 \}) 
\, {\uE}[d\widetilde{Q}^c_t]
= 
\int d\Gamma \frac{ \partial f(\{ {\bf q},{\bf p} \},t) }{\partial t}
H (\{ {\bf q},{\bf p} \}, \vec{\lambda}_t) 
 \, , \label{eqn:heat_SE}
\end{eqnarray}
represents the mean absorbed heat.
Here $d\Gamma$ denotes the phase space volume element for $\{ {\bf q},{\bf p} \}$.
These definitions lead to an equation analogous to the first law of thermodynamics,
\begin{eqnarray}
E^c_{t+dt} - E^c_{t} = dQ^c_t + dW^c_t \, . \label{eqn:1st}
\end{eqnarray}
Moreover, the Shannon information entropy \cite{jaynes}
\begin{eqnarray}
S^c(t) = -k_B \int d\Gamma \, f (\{ {\bf q},{\bf p} \},t) \ln f (\{ {\bf q},{\bf p} \},t) \label{eqn:S_ent}
\end{eqnarray}
enables us to deduce the following inequality:
\begin{eqnarray}
\lefteqn{\frac{dS^c(t)}{dt} 
- \sum_{i=1}^N  k_B \beta_i \frac{dQ^c_{(i)t}}{dt}} && \nonumber \\
&=&
\sum_{i=1}^N  k_B \int d\Gamma \frac{\beta_i}{f (\{ {\bf q},{\bf p} \},t)}
\left[
\gamma_{p_i} \sum_{\alpha=1}^D
\left( 
 \frac{1}{\beta_i} \frac{\partial f (\{ {\bf q},{\bf p} \},t)}{\partial p^\alpha_i} + \frac{\partial H(\{ {\bf q}_t , {\bf p}_t \},  \vec{\lambda}_t)}{\partial p^\alpha_i} f (\{ {\bf q},{\bf p} \},t)
\right)^2
 \right. \nonumber \\
&& \left. +
\gamma_{q_i} \sum_{\alpha=1}^D
\left( 
 \frac{1}{\beta_i}   \frac{\partial f (\{ {\bf q},{\bf p} \},t) }{\partial q^\alpha_i }+ \frac{\partial H(\{ {\bf q}_t , {\bf p}_t \},  \vec{\lambda}_t)}{\partial q^\alpha_i} f (\{ {\bf q},{\bf p} \},t)
\right)^2
 \right] \ge 0 \, .
\label{eqn:gene_clausius}
\end{eqnarray}
This is analogous to the Clausius inequality in a thermodynamical cycle 
if $\gamma_{q(i)}$ and $\gamma_{p(i)}$ are taken as time-dependent functions such that the system interacts only 
with heat baths with the same temperature at a time. 
When all temperatures of the heat baths are the same, 
$\beta_1 = \cdots = \beta_N = \beta$, 
the above inequality reduces to a continuous time-version of the second law:
\begin{eqnarray}
\frac{dS^c(t)}{dt}  - k_B \beta  \frac{dQ^c_t}{dt} \ge 0\, .
\end{eqnarray}

\section{Canonical quantization} \label{sec:cq}

Following the standard canonical quantization procedure \cite{sakuray}, 
we consider the self-adjoint operators associated with the classical variables 
$({\bf q}_{(i)}, {\bf p}_{(i)})$ satisfying
\begin{eqnarray} \label{eqn:boso_cr}
  [\hat{q}^\alpha_{(i)}, \hat{p}^\beta_{(j)}] &=& \ii \hbar \delta_{\alpha\beta}\delta_{ij} \, , \\
 {\protect  [\hat{q}^\alpha_{(i)}, \hat{q}^\beta_{(j)}] } &=&  0 \, , \\
{\protect [\hat{p}^\alpha_{(i)}, \hat{p}^\beta_{(j)}] } &=& 0 \, ,
\end{eqnarray}
where ``$\,\hat{\bullet}\,$''is introduced to denote operators.
We further admit the following quantization rules
\begin{eqnarray}
f(\{ {\bf q}, {\bf p} \},t) &\longrightarrow& \hat{\rho} (t)\, , \\
\left\{ g, h \right\}_{\text{PB}} &\longrightarrow& - \frac{\ii}{\hbar} [\hat{g}, \hat{h}] \, , \\
e^{\pm\beta_i H } g &\longrightarrow& e^{\pm \beta_i \hat{H}/2} \hat{g}e^{\pm \beta_i \hat{H}/2} \, ,
\end{eqnarray}
where $\hat{\rho}(t)$ is the density operator describing the system state.
The last rule is a key assumption in our approach, which is inspired by the relation satisfied 
between the time generators in Hamiltonian and stochastic dynamics considered 
in Refs.\ \cite{koide_rbm2}. See also the discussion in 
Refs. \cite{koide_fp1,koide_fp2}. 
A similar relation is, further, associated with detailed balance conditions in Ref.\ \cite{kurchan}.

Applying the above quantization rules to the generalized Kramers equation (\ref{eqn:kramers}), 
a master equation for the density operator is obtained,
\begin{eqnarray}
\frac{d}{dt}\hat{\rho} (t)
&=& 
\frac{\ii}{\hbar}\ [ \hat{\rho}(t), \hat{H} ]  \nonumber \\ 
&&-   \sum_{i=1}^{N} \sum_{\alpha=1}^D \frac{\gamma_{p_i} }{\beta_i \hbar^2}
\left [ e^{-\beta_i \hat{H}/2} 
\left[ e^{\beta_i \hat{H}/2} \hat{\rho}(t) e^{\beta_i \hat{H}/2}, \hat{q}^\alpha_{(i)} \right] e^{-\beta_i \hat{H}/2}, \hat{q}^\alpha_{(i)} \right] \nonumber \\
&& -   \sum_{i=1}^{N} \sum_{\alpha=1}^D   \frac{\gamma_{q_i} }{\beta_i \hbar^2} 
\left[ e^{-\beta_i \hat{H}/2} \left[ e^{\beta_i \hat{H}/2} \hat{\rho}(t)e^{\beta_i \hat{H}/2},  \hat{p}^\alpha_{(i)}\right]e^{-\beta \hat{H}/2}, \hat{p}^\alpha_{(i)} \right] \, , 
\label{eqn:qdd}
\end{eqnarray}
where $\hat{H}$ is a Hamiltonian operator. 
When all temperatures are the same, it is easily seen that the stationary solution is given by 
the thermal equilibrium state, $d e^{-\beta \hat{H}}/dt = 0$. 
The above equation can be mapped into the generalized form of the quantum 
Fokker-Planck equation considered in Ref.\ \cite{oliveira3}.

The evolution described by the master equation, however, does not necessarily satisfy the requirement of the CPTP map and thus we wonder under which circumstances this will happen. 
In the following, we will consider a set of harmonic oscillators interacting with heat baths in Eq.\ (\ref{eqn:qdd}) 
and show that this evolution will be reduced to the GKSL structure, 
satisfying automatically the detailed balance condition.

Writing the Hamiltonian of the system as 
\begin{eqnarray}\label{eqn:GKSL_ham}
\hat{H} 
=
\sum_{i=1}^{N}
\hat{H}_{(i)} 
=  \sum_{i=1}^{N} \sum_{\alpha=1}^D \frac{ 1}{2m_i} \hat{p}^{\alpha \,  2}_{(i)} + 
\frac{m_i\omega_i^2}{2} \hat{q}^{\alpha \,  2}_{(i)} \, ,
\end{eqnarray}
where $m_i$ and $\omega_i$ are masses and angular frequencies, respectively.
In general, $\omega_i$ is regarded as external parameters which control the width of the harmonic oscillator potentials and thus is time-dependent, as is considered in Refs.\ \cite{koide_fp1,koide_fp2}.
This time dependence however does not affect the following discussions and will be omitted for simplicity.
In this case, the master equation is reexpressed as a time-convolutionless form \cite{breuer2016}
\begin{eqnarray}\label{eqn:TCL_me}
\frac{d}{dt}\hat{\rho} (t)
=
\frac{\ii}{\hbar}\ [ \hat{\rho}(t), \hat{H}^\prime ]
+ \sum_{i=1}^{N} \sum_{\alpha=1}^D {\mathcal L}_{(i)}^\alpha [\hat \rho(t)], 
\end{eqnarray}
where the effective Hamiltonian is 
\begin{eqnarray} \label{eqn:TCL_ham}
\hat H^\prime &=& \hat H -  
\sum_{i=1}^{N} \sum_{\alpha=1}^D
 \frac{ \delta_i \gamma_{p_i}}{2\beta_i \hbar m_i\omega_i} 
\sinh(\beta_i \hbar \omega_i/2) [\hat{q}^{\alpha}_{(i)}, \hat{p}^{\alpha}_{(i)}]_+ \, ,\\
\delta_i &=& \frac{\gamma_{q_i}}{\gamma_{p_i}} (m_i\omega_i)^2 - 1 \, ,
\end{eqnarray}
and the dissipative term is 
\begin{eqnarray}
{\mathcal L}_{(i)}^\alpha [\hat \rho(t)] 
&=&
- \frac{1}{2\hbar} \sum_{\mu,\nu=1}^4 g_{\mu\nu} 
\left[ [\hat{L}^{\alpha \, \dagger}_{(i),\mu} \hat{L}^{\alpha}_{(i),\nu}, \hat{\rho} (t)]_+ 
-2 \hat{L}^{\alpha}_{(i),\mu} \hat{\rho} (t) \hat{L}^{\alpha \, \dagger}_{(i),\nu} \right] \, , \\
g_{\mu\nu} &=& {\rm Diag}(1,1,1,-1) \, .
\end{eqnarray}
The Lindblad operators are defined by
\begin{eqnarray}
&& \hat{L}^\alpha_{(i), 1} =
\Gamma_{(i),1} \, \hat a_{(i)}^{\alpha}
\, , \,\,\,\,\,\,\,
\hat{L}^\alpha_{(i), 2} =
\Gamma_{(i),2}  \, 
\hat a_{(i)}^{\alpha\, \dag}\, , \\
&& \hat{L}^\alpha_{(i), 3} = \sqrt{\delta_i} \, 
\hat{q}^{\alpha}_{(i)} \, , \,\,\,\,\,\,\, 
\hat{L}^\alpha_{(i), 4} = \frac{\sqrt{\delta_i}}{m_i\omega_i} \, \hat{p}^{\alpha}_{(i)} \, , 
\end{eqnarray}
with 
\begin{eqnarray}
\Gamma_{(i),j} = \sqrt{\frac{(\delta_i + 2) \gamma_{p_i}}{2\beta_i m_i\omega_i}\exp\left(  \frac{(-1)^{j+1} \beta_i \hbar \omega_i}{2} \right)}\, ,
\end{eqnarray}
and the lowering operator 
$ \hat a_{(i)}^{\alpha} := [\hat{q}^{\alpha}_{(i)} + i \hat{p}^{\alpha}_{(i)}/(m_i \omega_i)]\sqrt{m_i \omega_i /2\hbar}$.

Due to the negative matrix element $g_{44} = -1$, 
the master equation is reduced to the GKSL equation by choosing $\delta_i = 0, \forall i$,  
which implies $\hat{L}^\alpha_{(i), 3} = \hat{L}^\alpha_{(i), 4} = 0, \forall i$ 
and $\hat H' = \hat H$. 
It should be emphasized that the ratio of the dissipative coefficients satisfies the detailed balance condition: 
$\Gamma_{(i),1}/\Gamma_{(i),2}=e^{\beta_i \hbar \omega_i/2}$.
Normally this condition is introduced by hand to describe thermal relaxation processes, but is induced automatically in the present approach.

As is well-known, in this case, the inequality analogous to the second law of thermodynamics can be introduced \cite{spohn,alicki,davies,spohn2,pusz}.
The von Neumann entropy, $S^q(t) = -k_B {\rm Tr} \left[ \hat{\rho} \ln \hat{\rho}\right]$, satisfies
\begin{eqnarray}\label{eqn:EntProd}
\frac{dS^q (t)}{dt} - \sum_{i=1}^{N} k_B \beta_i \frac{dQ^{q(i)}_{t}}{dt} \ge 0 \, ,
\end{eqnarray}
where the heat currents associated with the bath of temperature $\beta_i$ are defined by
\begin{eqnarray}
\frac{dQ^{q(i)}_{t}}{dt} = {\rm Tr} \left[\frac{d\hat{\rho}(t)}{dt} \hat{H}_{(i)} \right] \, . \label{eqn:heat_GKSL}
\end{eqnarray}
Comparing this with Eq.\ (\ref{eqn:heat_SE}), 
one can easily see that the representations of classical and quantum heats 
have a clear correspondence.
The introduction of the first law is straightforward by the identification 
$E^q_t = {\rm Tr} [ \hat{\rho}(t) \hat{H}]$.

It is noteworthy that, in the limit of vanishing $\gamma_{q_i}$, or $\delta_i = -1$, 
the generalized Kramers equation (\ref{eqn:kramers}) is reduced to its standard version, 
as explained below Eq.\ (\ref{eqn:eq_p}). 
In this situation, it is easy to see that $\hat{L}^\alpha_{(i), 4}$ does not disappear and the 
master equation is not a GKSL equation.

The employed definitions of thermodynamical quantities are still applicable to the generic time-convolutionless master equation (\ref{eqn:TCL_me}), 
but the positivity of the entropy production expressed in Eq.\ (\ref{eqn:EntProd}) is not necessarily satisfied.
Indeed this positivity is associated with Markovian properties of the master equation \cite{colla}. 
We will return to this point in the concluding remarks.

A remarkable consequence of the algebraic structure of the mater equation (\ref{eqn:qdd}) is that 
our formulation is applicable not only to Bosons but also to Fermions. 
The fermionic excitation of an harmonic oscillator is described by the anti-commutation relations 
$[ \widehat{p} , \widehat{q} ]_+ = 0$,   
$[\hat{p},\hat{p}]_+ /(m\omega)= m\omega [\hat{q},\hat{q}]_+ = \hbar/2$
and the Hamiltonian operator $\widehat{H} = \ii \omega \widehat{q} \widehat{p}$. 
The raising and lowering operators are defined in the same fashion as the bosonic harmonic oscillator 
and then satisfy fermionic anti-commutation relations. 
Applying these to the master equation (\ref{eqn:qdd}) and choosing parameters appropriately, 
the GKSL equation for fermions is obtained. The heat and the entropy of this fermionic system are shown to satisfy the inequality analogous the second law again.

\section{concluding remarks} \label{sec:conc}

We introduced a generalized classical model for describing thermal relaxation processes 
by considering the interaction with heat baths even through the equation of velocity.
In this  model, we can still 
define laws analogous to the thermodynamical ones.
Applying the canonical quantization to this, a quantum master equation for the density operator is obtained.
This equation has a thermal equilibrium state as the stationary solution, but the time evolution is not necessarily 
the CPTP map. 
In the application to the harmonic oscillator potential, however, 
the requirement for the CPTP map is shown to be satisfied 
by choosing parameters appropriately and then our equation reproduces the GKSL equation with the detailed balance condition.

L. Onsager assumed that the average regression of thermal fluctuations behaves like the corresponding macroscopic irreversible process \cite{onsager}. 
This hypothesis is, however, considered not to be applicable to quantum systems \cite{grabert,talkner,ford,guarnieri,cosaccchi}.
It is worth investigating this problem using the generalized classical model proposed here.

The interaction with the environment described in our quantum master equation (\ref{eqn:qdd}) is not local 
due to potential inter-particle interactions present within the Hamiltonian $\hat{H}$ in its dissipative part,
{\it i.e.}, the thermal bath at temperature $T_i$ interacts with the $j$-th particle 
due to inter-particles interactions. 
This non-locality stems from the generalized Kramers equation (\ref{eqn:kramers}), 
which finds its roots in the nature of the friction forces, elucidated below Eqs. (\ref{eqn:eq_x}) and (\ref{eqn:eq_p}).
The non-locality is a significant aspect in investigating heat conduction within network models \cite{kosloff,volovich,cattaneo,basharov}.
There is an ongoing debate regarding whether the violation of the second law of thermodynamics 
is related to employing a local master equation.
In this regard, it is worth mentioning the research on heat conduction in stochastic energetics, for instance, in the chapter 4 of Ref.\ \cite{sekimoto}. 
These studies have successfully replicated the Fourier law and reproduced the second law 
when $\gamma_{q_i} = 0$, that is, in the standard Brownian motion. 
The issue of heat conduction within generalized Brownian motion ($\gamma_{q_i} \neq 0$) is however an open question. 
It might be possible to comprehensively understand the heat conduction 
in both classical and quantum systems by analyzing our quantum master equation.

There is a complementary relation between our quantum master equation and the GKSL equation.
In the GKSL equation, the time evolution is always a CPTP map but the stationary solution is not necessarily 
given by the thermal equilibrium state.
This is satisfied by introducing the detailed balance condition \cite{toscano}.
Contrary to this, our equation always has the equilibrium state as the stationary solution but the evolution 
is not necessarily given by a CPTP map. 
The compatibility of our master equation with a CPTP map for general Hamiltonian has not yet been known 
and this is the reason why we did not consider the inter-particle couplings in Sec.\ \ref{sec:cq}, 
even when the corresponding classical model with inter-particle couplings is shown to be 
consistent with thermodynamics by Eqs. (\ref{eqn:1st}) and (\ref{eqn:gene_clausius}).
The time evolution of the GKSL equation is a CPTP map, but not vice versa. Indeed there are non-Markovian master equations 
which satisfy the requirement of the CPTP map \cite{breuer2016}.  
Thus the applicability of our strategy to more general cases is an open question.

The interaction with environments (heat baths) plays an important role in understanding the quantum-classical transition in terms of decoherence \cite{schlosshauer}. 
That is, such an interaction has been considered crucial to distinguish classical and quantum worlds.
Meanwhile, our result suggests that the interactions with environment in classical and quantum worlds have a correspondence, which is a new perspective and will provide a new insight to the study of the measurement theory.

\vspace*{1cm}
The authors acknowledge P.\ Talkner for valuable comment.
T. K. acknowledges the financial supports by CNPq (No.\ 305654/2021-7). 
A part of this work has been done under the project INCT-Nuclear Physics and Applications (No.\ 464898/2014-5); 
F. N. is a member of the Brazilian National Institute of Science and Technology
for Quantum Information [CNPq INCT-IQ (465469/2014-0)].

\appendix

\section{Derivation of generalized Kramers equation} \label{app:0}

The normalized phase space distribution is defined by 
\begin{eqnarray}
f(\{{\bf q},{\bf p}\},t) = \int d\Gamma_0 \, f_0 (\{{\bf q}_0,{\bf p}_0\}) \, 
{\rm E}\!\left[  \prod_{i=1}^N \delta^{(D)}({\bf q} - \widetilde{\bf q}_{(i)t}) \delta^{(D)}({\bf p} - \widetilde{\bf p}_{(i)t})\right] \, ,
\label{eqn:psd}
\end{eqnarray}
where $\{ {\bf q}_0, {\bf p}_0 \}$ are the position and momentum at an initial time,  
$d\Gamma_0$ is the corresponding phase space volume, 
and $f_{0} (\{ {\bf q}_0, {\bf p}_0 \})$ is the initial probability distribution.

The generalized Kramers equation in Eq.\ (\ref{eqn:kramers}), 
governing the evolution of $f(\{{\bf q},{\bf p}\},t)$, is obtained performing the temporal derivative of (\ref{eqn:psd}). 
To this end, we will employ Ito's lemma \cite{gardiner}, 
a Taylor expansion for stochastic functions.  
In an example, when a stochastic variable $\tilde{x}_t$ satisfies 
\begin{eqnarray}
d\widetilde{x}_t = u(\widetilde{x}_t,t) dt + \sqrt{2\nu} d\widetilde{B}_t \, , \label{eqn:ex-x}
\end{eqnarray}
for a generic continuous function $u(x,t)$, a Wiener process $\tilde{B}_t$, and a positive constant $\nu$, 
the differential
of an arbitrary function of $\widetilde{x}_t$ is given by 
\begin{eqnarray}
d g(\widetilde{x}_t) = dt \left. \left[ u(\widetilde{x}_t,t) \partial_x  + \nu \partial^2_x \right] g(x) \right|_{x=\widetilde{x}_t} + o(dt) \, .
\end{eqnarray}

Now, differentiating Eq.\ (\ref{eqn:psd}) with respect to time through Ito's lemma 
and using Eqs.\ (\ref{eqn:eq_x}) and (\ref{eqn:eq_p}) instead of Eq.\ (\ref{eqn:ex-x}), 
we find 
\begin{eqnarray}\label{eq:apkram}
\partial_t f (\{{\bf q},{\bf p}\},t) 
= 
- \sum_{i=1}^N \left\{  \frac{\partial}{\partial {\bf q}(i)} \cdot {\bf J}_{{\bf q}(i)} +\frac{\partial}{\partial {\bf p}(i)} \cdot {\bf J}_{{\bf q}(i)} \right\} \, ,
\end{eqnarray}
where
\begin{align}
 {\bf J}_{{\bf q}(i)}
&=  
\left( \frac{\partial H}{\partial {\bf p}_{(i)}} - \gamma_{q_{i}} \frac{\partial H}{\partial {\bf q}_{(i)}} \right) 
- \frac{\gamma_{q_i}}{\beta_i}\frac{\partial}{\partial {\bf q}_{(i)}}  f (\{{\bf q},{\bf p}\},t)  \, ,\\
 {\bf J}_{{\bf p}(i)}
&=  
- \left( \frac{\partial H}{\partial {\bf q}_{(i)}} - \gamma_{p_{i}} \frac{\partial H}{\partial {\bf p}_{(i)}} \right) 
- \frac{\gamma_{p_i}}{\beta_i}\frac{\partial}{\partial {\bf p}_{(i)}}  f (\{{\bf q},{\bf p}\},t)  \, .
\end{align}
Finally, using the Poisson bracket in Eq.\ (\ref{eqn:pb}) to rewrite the right-hand side of (\ref{eq:apkram})   
and noting that $\beta_i e^{\beta_i H} \{ H, G \}_{\rm PB} = \{ e^{\beta_i H}, G \}_{\rm PB} $, 
Eq.\ (\ref{eqn:kramers}) is obtained.

\section{Asymptotic behavior} \label{app:1}

When all temperatures of the heat baths are the same $\beta_1= \cdots=\beta_n = \beta$ and the external parameters are constants 
$\vec{\lambda}_t = \vec{\lambda}$, 
the stationary solution of the generalized Kramers equation is given by the thermal equilibrium distribution, 
\begin{eqnarray}
f^*(\{ {\bf q},{\bf p} \}) = \frac{1}{Z} e^{-\beta H (\{ {\bf q}_t , {\bf p}_t \},  \vec{\lambda})}  \, ,
\end{eqnarray}
where $Z$ is a normalization constant,
\begin{eqnarray}
Z = \int d\Gamma \, f^* (\{ {\bf q},{\bf p} \}) \, .
\end{eqnarray}
Indeed, we can define the Kullback-Leibler divergence by
\begin{eqnarray}
J( f | f^* ) 
= \int d\Gamma f  \ln \frac{f }{f^* } \, .
\end{eqnarray}
This is a monotonically decreasing function and asymptotically converges to $f^*$, 
\begin{eqnarray}
\frac{d}{dt}J( f | f^*) \le 0 \, .
\end{eqnarray}

\section{Generalization of concept of work} \label{app:2}

We consider a single particle of mass $m$ described by 
\begin{eqnarray}
\frac{d{\bf q}}{dt} &=& \frac{\bf p}{m} + {\bf F}^{(q)}_{ex} \, \\
\frac{d{\bf p}}{dt} &=& - \nabla V + {\bf F}^{(p)}_{ex} \, 
\end{eqnarray}
Here ${\bf F}^{(p)}_{ex}$ is an external force.
We further assume that the relation between the velocity and momentum in the first equation is modified by an external perturbation ${\bf F}^{(q)}_{ex}$. 
As pointed out in the introduction, such a modification is observed in the stochastic formulation of quantum mechanics \cite{nelson,yasue,koide_review}.
Then the change of the energy of the particle, $E ={\bf  p}^2/(2m) + V$, is represented by 
\begin{eqnarray}
\frac{d}{dt} E
=
\frac{{\bf p}}{m} \cdot \frac{d{\bf p}}{dt} + \nabla V \cdot \frac{d{\bf q}}{dt} 
=
 {\bf F}^{(p)}_{ex} \cdot \frac{d{\bf q}}{dt} 
- \frac{d{\bf p}}{dt} \cdot {\bf F}^{(q)}_{ex} \, .
\end{eqnarray}
The first term on the right-hand side is the standard work and the second term is a new contribution. 
This result is used to define the heat in the present model.

\end{document}